\documentclass[11pt]{article}
\usepackage{moriond,epsfig,xspace,dcolumn}
\input babarsym
\newcommand{\etaCP}{\ensuremath{\eta_{\CP}}}

\def\Dp      {\ensuremath{D^+}}
\def\Dm      {\ensuremath{D^-}}
\def\Dstar   {\ensuremath{D^{*}}}
\def\Dstarp  {\ensuremath{D^{*+}}}
\def\Dstarm  {\ensuremath{D^{*-}}}

\def\PRL{{\em Phys. Rev. Lett.}}
\def\PRD{{\em Phys. Rev.} D}

\def\ra{\rightarrow}

\def\be{\begin{equation}}
\def\ee{\end{equation}}
\def\bea{\begin{eqnarray}}
\def\eea{\end{eqnarray}}

\providecommand{\hips}{\mbox{\ensuremath{\ps^{-1}}}}
\providecommand{\Brec}{\mbox{\ensuremath{B_{\rm rec}}}}
\providecommand{\Btag}{\mbox{\ensuremath{B_{\rm tag}}}}

\def\DorDstarm {\ensuremath{D^{(*)-}}\xspace}
\def\DorDstarzb  {\ensuremath{\Dbar^{(*)0}}\xspace}
\newcommand{\deltaE} {\ensuremath{\Delta \rm{E}}\xspace}

\newcommand{\Bflav}   {\ensuremath{\B_{\rm{flav}}}\xspace}
\newcommand{\Bcp}   {\ensuremath{\B_{\CP}}\xspace}

\begin{document}
\vspace*{4cm}
\title{Measurements of Lifetimes, Mixing and \CP\ Violation 
of B Mesons with the \babar\ Detector}

\author{ G. RAVEN \\ (for the \babar\ collaboration)}

%\address{Department of Physics, Theoretical Physics, 1 Keble Road,\\
%Oxford OX1 3NP, England}
\address{Department of Physics, 
University of California, San Diego\\
9500 Gilman Drive, La Jolla, CA 92093 }

\maketitle\abstracts{
Using a data sample of 62 million $\FourS\to B\Bbar$
decays collected between 1999 and 2001 by the \babar\ detector at the
\pep2\ asymmetric-energy \BF\ at SLAC we study events
in which one neutral $B$ meson is fully reconstructed in a final state
containing a charmonium meson and the flavour of the other neutral $B$
meson is determined from its decay products. The amplitude of the
\CP-violating asymmetry, which in the Standard Model is proportional to
\stwob, is derived from the decay time distributions. We
measure $\stwob = 0.75 \pm 0.09\ \stat \pm 0.04\ \syst$ and
$|\lambda| = 0.92 \pm 0.06\ \stat \pm 0.02\ \syst$. The latter is
consistent with the Standard Model expectation of no direct 
\CP\ violation. These results are preliminary. In addition, we 
report on precision measurements of the $B$ lifetimes, and the 
\Bz-\Bzb\ oscillation frequency \deltamd.
}

\section{Introduction}

\CP\ violation has been a central concern of particle physics since its
discovery in 1964 in the decays of \KL\ decays\cite{EpsilonK}.
An elegant explanation of the
\CP-violating effects in these decays is provided by
the \CP-violating phase of the three-generation Cabibbo-Kobayashi-Maskawa
(CKM) quark-mixing matrix~\cite{CKM}.
However, existing studies of \CP\ violation
in neutral kaon decays and the resulting experimental constraints on the
parameters of the CKM matrix~\cite{MSConstraints} do not provide a stringent
test of whether the CKM phase describes \CP\ violation~\cite{Primer}.
In the CKM picture, large \CP\ violating asymmetries are expected in
the time distributions of \Bz\ decays to charmonium final states. 

In general, \CP\ violating asymmetries are due to the interference
between amplitudes with a weak phase difference. 
For example,
a state initially produced as a \Bz\ (\Bzb) can decay to a \CP\ eigenstate
such as $\jpsi \KS$ directly
or can oscillate into a \Bzb\ (\Bz) and then decay
to $\jpsi \KS$.  With little theoretical uncertainty in the Standard Model, the 
phase difference between these amplitudes is equal to twice the angle
$\beta = \arg \left[\, -V_{\rm cd}^{ }V_{\rm cb}^* / V_{\rm td}^{ }V_{\rm tb}^*\, \right]$ of the
Unitarity Triangle.  The \CP-violating asymmetry 
in this mode allows a direct determination of \stwob, and
can thus provide a crucial test of the Standard Model.

A \BzBzb\ pair produced in \FourS\ decays
evolves in a coherent $P$-wave until one of the \B\ mesons decays.
If one of the \B\ mesons, referred to as \Btag, can be 
ascertained to decay to a state of known flavour, {\em i.e.} \Bz\ or
\Bzb, at a certain 
time $t_{\rm tag}$, 
the other \B, referred to as \Brec, {\it at that time} must be of the
opposite flavour as a consequence of Bose symmetry.
Consequently, the oscillatory probabilities for observing
\BzBzb, $\Bz\Bz$ and $\Bzb\Bzb$ pairs produced in
\FourS\ decays are a function of
$\deltat = t_{\rm rec} - t_{\rm tag}$, allowing
mixing frequency and \CP\ asymmetries to be determined
if \deltat\ is known.

At the PEP-II asymmetric $e^+e^-$ collider~\cite{pepii}, resonant production of
the \FourS\  provides a
copious source of $\BzBzb$ pairs moving along the beam axis ($z$ direction)
with an average Lorentz boost of $\left<\beta\gamma\right> = 0.55$.
Therefore, the proper decay-time difference $\deltat$ is,
to an excellent approximation, proportional to
the distance \deltaz\ between  the two \Bz-decay vertices
along the axis of the boost,
$\deltat \approx \deltaz / c\left< \beta \gamma \right>$.
The average separation between the two $B$ decay vertices is
$\deltaz \approx \left<\beta\gamma\right> c \tau_B = 260$\mum, while the RMS
$\deltaz$ resolution of the detector is about
180\mum.

The lifetime, mixing and \stwob\ analyses share a common analysis strategy:
\begin{itemize}
  \item select events where one \B, labeled \Brec, is fully reconstructed;
  \vspace*{-0.1cm}
  \item determine the vertex of the other $B$ decay, \Btag, in the event
        by performing a vertex fit to the remaining charged particle trajectories, and compute \deltat.
\end{itemize}
At this point, one can perform an unbinned likelihood fit to the
\deltat\ distribution and determine the $B$ lifetime. The mixing and \CP\ 
asymmetry measurements require one additional step:
\begin{itemize}
  \item determine the flavour of the \Btag\ decay.
\end{itemize}
To determine the oscillation frequency \deltamd\, events are selected where \Brec\ is reconstructed
in a neutral decay mode with a known flavour, such as $\Dstarp\pim$, and a simultaneous unbinned 
likelihood fit to the \deltat\ distributions of events where
\Brec\ and \Btag\ have opposite flavour (unmixed) and equal flavour (mixed)
is performed.
To determine \CP\ asymmetries, \Brec\ is a reconstructed 
\CP\ final state such as $\jpsi\KS$, and the \deltat\ distributions
of events where \Btag\ is a \Bz\ and \Bzb\ respectively are fitted 
simultaneously.

In order to establish the experimental technique, we first 
present precision measurements of the \Bz\ and \Bu\ lifetimes, and the
\Bz\Bzb\ oscillation frequency \deltamd. These measurements share the
same vertexing algorithm and \deltat\ determination 
as the measurement of the \CP\ asymmetries, and, in case of the mixing
measurement, the same flavour tagging algorithm. 
To eliminate possible experimenter's bias the parameter under study was hidden 
in all analyses until the event selection and reconstruction, fitting 
procedure and systematic errors were finalized.

\section{The \babar\ detector and data sets}

The data used were recorded with the \babar\ detector
in the period October 1999--December 2001.
The total integrated luminosity of the data set is equivalent to
56\invfb\ collected near the \FourS\ resonance.
% and XX\invfb\ collected 40\mev\ below the \FourS\ resonance
 The corresponding number
of produced \BB\ pairs is estimated to be about 62 million. 
The measurement of the charged and neutral \B\ lifetimes is based on the
initial 23 million \BB\ pairs, whereas the dataset used for the \deltamd\ measurement 
includes the first 32 million \BB\ pairs.  This latter sample was also
used for the previously published measurement\cite{sin2b-prl} of \stwob;
in contrast, the preliminary measurement of \stwob\ described here 
utilizes the entire data sample available. %% FIXME: mention improved
% alignments on 1999-2000 sample here?

Since the \babar\ detector is described in detail
elsewhere~\cite{babar-detector-nim},
only a brief description is given here.
Surrounding the beam-pipe is a 5-layer silicon vertex tracker (SVT), which
provides precise measurements of the trajectories of charged particles
as they leave the \epem interaction point.
Outside of the SVT, a 40-layer drift chamber (DCH)
allows measurements of track momenta in a 1.5\,T magnetic field as well as
energy-loss measurements, which contribute to charged
particle identification. Surrounding the DCH is a detector of internally
reflected Cherenkov radiation (DIRC), which provides charged hadron
identification. Outside of the DIRC is a CsI(Tl) electromagnetic
calorimeter (EMC) that is used to detect photons, provide electron
identification and reconstruct neutral hadrons. The EMC is surrounded by a
superconducting coil, which creates the magnetic field for momentum
measurements.  Outside of the coil, the
flux return is instrumented with resistive plate chambers interspersed with
iron (IFR) for the identification of muons and long-lived neutral hadrons.
We use the GEANT4 package~\cite{geant4} to simulate interactions of particles
traversing the \babar\ detector.

\section{Measurement of \B\ Lifetimes and \deltamd}

\subsection{Exclusive \B\ reconstruction}

The so-called \Bflav\ sample used for lifetime and mixing
analyses consists of events where \Brec\ is reconstructed
in the modes $\Bz \ra \DorDstarm \pip$, $\DorDstarm \rho^+$, $\DorDstarm
a_1^+$, $\jpsi  \Kstarz$ and $\Bu \ra \DorDstarzb \pip$, $\jpsi K^+$,
$\psitwos K^+ $.
Charged and neutral \Dstarb\ candidates are formed by combining a \Dzb
with a \pim\ or \piz . \Dzb\ candidates are reconstructed in the decay
channels $\Kp\pim$, $\Kp\pim\piz$, $\Kp\pip\pim\pim$ and $\KS \pip\pim$
and \Dm\ candidates in the decay channels $\Kp \pim \pim$ and $\KS\pim$.
We reconstruct \jpsi and \psitwos in the decays to $\epem$ and
$\mu^+\mu^-$ and the \psitwos decay to \jpsi \pip \pim .

\begin{figure}[htbp]
\caption{Beam energy substituted mass distribution for selected 
 \Bz\ (a) and \Bu\ (b) candidates. In 56 \invfb, we reconstruct 15K
  \Bz and 13K \Bp  flavour tagged signal events. Average signal purities for
  $\mes>5.27\gevcc$, are 85~\% and  89~\%, respectively. 
\label{fig:hadronic}
}
\begin{center}
\mbox{
\includegraphics[width=0.45\textwidth]{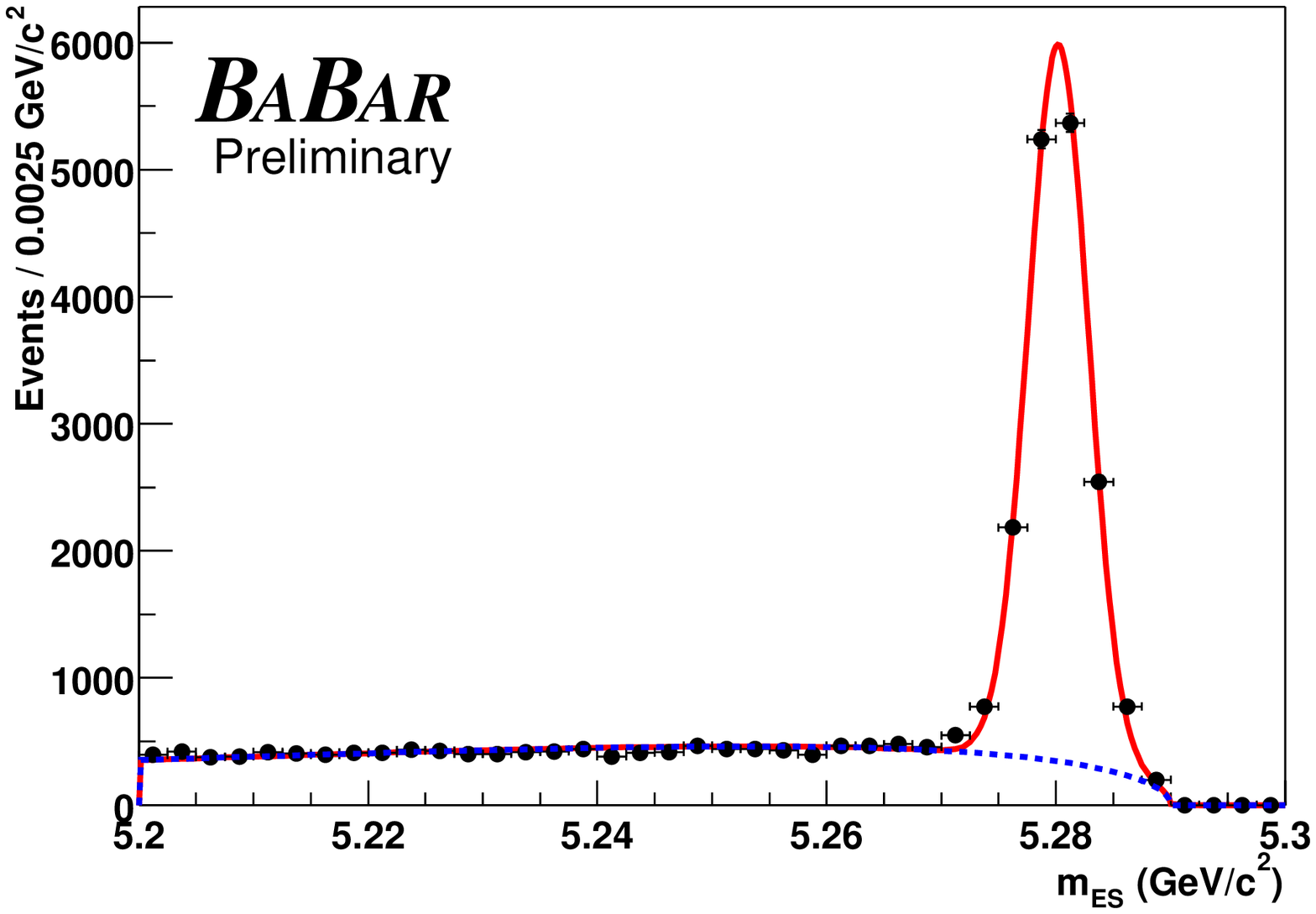}
\includegraphics[width=0.45\textwidth]{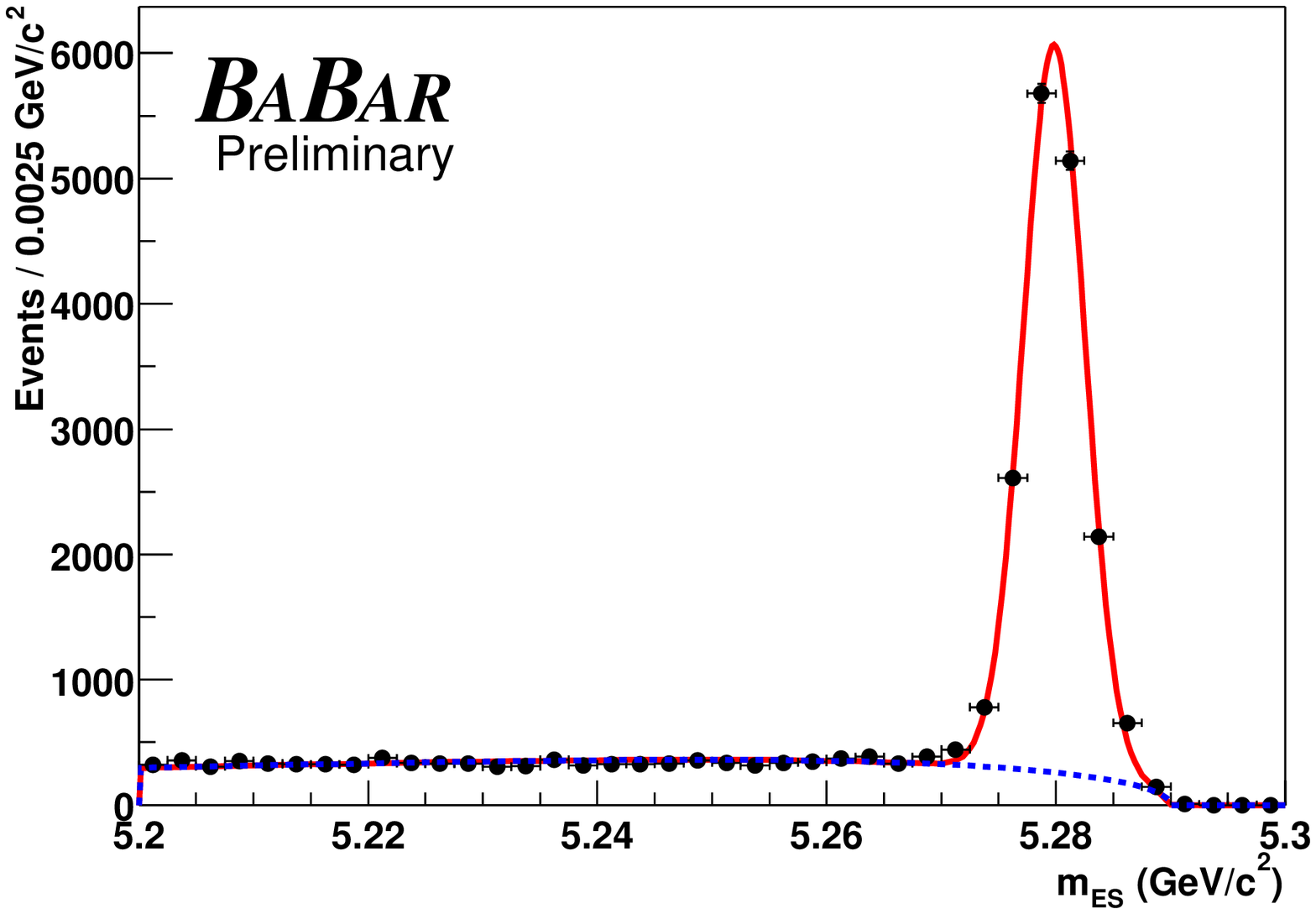}
\put(-250,115){{\large a)}}
\put(-40,115){{\large b)}}
}
\end{center}
\end{figure}

Continuum $e^+e^-\to q\overline q$ background is suppressed by
requirements on the normalized second Fox-Wolfram moment~\cite{fox}
for the event and on the angle between the thrust axes of \Brec\ and of
the other $B$ in the event. \Brec\ candidates are identified by
the difference \deltaE between the reconstructed $B$ energy
and $\sqrt{s}/2$ in the \FourS\ frame, and the
beam-energy substituted mass
\mes\ calculated from $\sqrt{s}/2$ and the reconstructed $B$
momentum. We require $\mes > 5.2$\gevcc
and $|\deltaE| < 3 \sigma_{\deltaE}$. The distributions of \mes\ for
selected \Bflav\ candidates is shown in Fig.~\ref{fig:hadronic}.

\subsection{\deltat\ determination}

The decay time difference, \deltat,
between the two \B\ decays is determined from the
measured separation, $\Delta z = z_{rec}-z_{tag}$,
along the $z$ axis between the reconstructed $\Brec (z_{rec})$ and
flavour-tagging decay \Btag\ vertex $(z_{tag})$.
This measured $\Delta z$ is converted into \deltat with the
known 
\FourS\ boost, including a
correction on an event-by-event basis for the direction of the \B\ mesons
with respect to the $z$ direction in the \FourS\ frame.
The \deltat\ resolution is limited by the $z$ resolution of the tagging vertex.
The \Btag\ decay vertex reconstruction starts from all
tracks in the event except those incorporated in $B_{rec}$. 
%In order to
%reduce the contamination from $D$ meson decay products, those identified as kaons
%are also excluded.
An additional constraint is provided by the calculated
\Btag\ production point and three-momentum, determined
%from the
%three 
%momentum of the $B_{rec}$ candidate, its decay vertex,
%and the average position of the interaction point and the \FourS\ boost.
%This is determined 
from the three-momentum
of the fully reconstructed \Brec\ candidate,
its decay vertex, and
the average position of the interaction point (with a vertical size of 10\mum) 
and the \FourS\ average boost.
The derived \Btag\ trajectory is fit to a common vertex with the
remaining tracks in the event. % (excluding those from $B_{rec}$).
Reconstructed $\KS$ or $\Lambda$
candidates are used as input to the fit in place
of their daughters in order to reduce
bias due to long-lived particles.
Tracks with a large contribution to the $\chi^2$
are iteratively removed from the fit
until all remaining tracks have a reasonable fit probability
or all tracks are removed. For 99.5\% of the reconstructed events the r.m.s 
\deltaz\ resolution is 180 \mum.
%Only events for which $|\deltat|<20$\ps and $\sigma_{\deltat} < 2.5 (1.4)$\ps are
%retained in the \stwob\ (\deltamd) analysis,
%where $\sigma_{\deltat}$ is the measurement error derived from the
%vertex fits.

Two different parameterizations are used to model the decay-time
difference resolution. In the measurements of \deltamd\ and \stwob,
the time resolution function
is approximated by a sum of three Gaussian distributions 
(core, tail, and outlier) with different means and widths,
\begin{equation}
{\cal {R}}( \delta_{\rm t},\sigma_{\deltat} ; \hat {a} = \{ f_k,S_k,b_k,\sigma_3 \} ) =  \sum_{k=1}^{2} 
{ \frac{f_k}{S_k\sigma_{\deltat}\sqrt{2\pi}} \, {\rm exp} 
\left(  - \frac{( \delta_{\rm t}-b_k\sigma_{\deltat})^2}{ 
 2({S_k\sigma_{\deltat}})^2 }  \right) } + \nonumber 
 { \frac{f_3}{\sigma_3\sqrt{2\pi}} \, {\rm exp} 
\left(  - \frac{ \delta_{\rm t}^2}{ 
 2{\sigma_3}^2 }  \right) },\nonumber
\label{eq:vtxresolfunct}
\end{equation}
%\begin{equation*}
%{\cal R}( \deltat ; \hat {a}_i ) =  \sum_{k=1}^{3}
%{ \frac{f_k}{\sigma_k\sqrt{2\pi}} {\rm e}^{
%-(\deltat-\deltat_{true}-\delta_{k,i} )^2/2{\sigma_k}^2} }.
%\end{equation*}
where $\delta_{\rm t}=\deltat-\deltat_{\rm true}$. 
For the core and tail Gaussians, the widths
$\sigma_{k}=S_{k}\times\sigma_{\deltat}$ are the event-by-event
measurement errors scaled by a common factor $S_{k}$.
The scale factor of the tail Gaussian is fixed to the Monte Carlo value
since it is strongly correlated with the other resolution function
parameters.
The third Gaussian, with a fixed width of $\sigma_3=8$\ps, accounts for
outlier events with incorrectly reconstructed vertices (less than 1\% of events).
%A separate core bias coefficient, $b_{1,i}$, is allowed
%for each tagging category $i$ to account for small
%shifts due to inclusion of charm decay products in the tag vertex,
%while a common bias coefficient, $b_2$, is used for the
%tail component. 
The offsets $b_i$ are modeled to be proportional
to $\sigma_{\deltat}$, which is correlated with the weight that the
remaining daughters of charm particles have in the tag vertex reconstruction.
The tail and outlier fractions and the scale factors
are assumed to be the same for all decay modes,
since the precision of the \Btag\ vertex measurement is the limiting
factor for the \deltat\ resolution. This
assumption is confirmed by Monte Carlo studies.

The three Gaussian resolution function is less suited for the measurements
of \B\ lifetimes due to the large correlation between the resolution function
parameters and the lifetimes, which leads to increased statistical errors.
Studies with Monte Carlo simulations and data show that the sum of a 
zero-mean Gaussian distribution and its convolution with a one-sided 
exponential provides a good trade-off between statistical and 
systematic uncertainties in the lifetime measurement:

\begin{eqnarray}
\label{eq:resol}
{\cal R} (\delta_{t} , \sigma_{\deltat} | \hat{a}= \{ f_1,s,\kappa \}) = 
f_1    \frac{1}{\sqrt{2\pi}  s \sigma_{\deltat}}   \exp\left(
-\frac{\delta_{t}^2}{2s^2 \sigma_{\deltat}^2}
\right) \ \ \ \ \ \ \ \ \ \ \ \ \ \\
 + \int_{-\infty}^{0} \frac{1-f_1}{\kappa\sigma_{\deltat}}  \exp\left(
\frac{\delta_{t}'}{\kappa\sigma_{\deltat}} \right)   
\frac{1}{\sqrt{2\pi}  s \sigma_{\deltat}}   \exp\left(
-\frac{(\delta_{t} - \delta_{t}')^2}{2s^2 \sigma_{\deltat}^2}\right)
{\rm d}(\delta_{t}')\; .\nonumber 
\end{eqnarray}
The parameters $\hat{a}$ are  
the fraction $f$ in the core Gaussian component, 
a scale factor $s$ for the per-event errors $\sigma_{\deltat}$, and the 
factor~$\kappa$ in the effective time constant~$\kappa\sigma_{\deltat}$
of the exponential which accounts for charm decays. $\Delta t$ 
  outlier events are modeled the same way as in the three Gaussian
resolution function. 
%
%The resolution function parameters are assumed to be the same for all
%\Bz and \Bp decay modes. % This assumption is confirmed by Monte Carlo simulation studies. 
The resolution functions differ only slightly
between \Bz and \Bu mesons due to different mixtures of $D^-$ and \Dzb\
mesons in the  \Btag\ decays and we use a single set of resolution
function parameters for both \Bz\ and~\Bu\ in the lifetime fits.

\subsection{Lifetime results}
We extract the \Bp and \Bz lifetimes from an unbinned maximum likelihood
fit to the \deltat distributions of the selected $B$ candidates. The
probability for an event to be signal is estimated from $\mes$~fits  
(Fig.~\ref{fig:hadronic}) and the \mes\ value of the \Brec\ candidate.
In the likelihood, the probability density for the signal events is
given by  
\begin{equation}
\label{eq:Phi}
{\cal G}(\deltat,\sigma_{\deltat} | \tau,\hat{a}) = \int_{-\infty}^{+\infty}
e^{-|\deltat |/\tau}/(2\tau)
%g(\deltat' | \tau) 
\, \calR \it (\deltat-\deltat', \sigma_{\deltat} |\hat{a}) \; {\rm d}(\deltat'),
\end{equation}
and the background \deltat\ distribution for each \B\ species
is empirically modeled by the sum of a prompt component and a lifetime
component convolved with the same resolution function, but with a
separate set of parameters. 
The likelihood fit involves 17 free parameters in addition to the \Bz
and the \Bp lifetimes: 12 to describe the background \deltat
distributions and 5 for the signal resolution function.
The charged $B$ lifetime $\tau_{\Bu }$ is replaced with $\tau_{\Bu } =
r \cdot \tau_{\Bz }$ to estimate the statistical error on the
ratio~$r= \tau_{\Bu}/\tau_{\Bz}$.  

We determine the \Bz and \Bu meson lifetimes and their ratio to be:%\\[-8mm]
\begin{eqnarray}
 \tau_{\Bz} &=& 1.546 \pm 0.032\mbox{ \stat } \pm 0.022\mbox{ \syst} \mbox{ ps,}\nonumber \\
 \tau_{\Bu} &=& 1.673 \pm 0.032\mbox{ \stat } \pm 0.023\mbox{ \syst} \mbox{ ps, and}\nonumber \\
 \tau_{\Bu}/\tau_{\Bz} &=& 1.082 \pm 0.026\mbox{ \stat} \pm 0.012\mbox{ \syst}.\nonumber 
\end{eqnarray}

These are the most precise published measurements to date~\cite{blife-prl} and
are consistent with the world averages~\cite{PDG2000}. The resolution
function parameters are consistent with those found in a Monte Carlo
simulation that includes detector alignment effects. 
%Figure~\ref{fig:dt-life-n-mix} shows the results of the likelihood fit 
%superimposed on the \deltat~distributions for \Bz\ and \Bu\ events. 
With the current data sample these measurements are still statistically 
limited. The dominant systematic errors arise from uncertainties in the 
description of the combinatorial background and of events with large
\deltat\ values, the use of a common time resolution function for $B^0$
and $B^+$ and from limited Monte Carlo statistics.

\subsection{Flavour tagging}

After the daughter tracks of the \Brec\ are removed from the event,
the remaining tracks are analyzed to determine the flavour of
the \Btag, and this ensemble is assigned a tag flavour, either \Bz\ or \Bzb.
For this purpose, flavour tagging information carried by
primary leptons from semileptonic $B$ decays, charged kaons, soft pions from
\Dstar\ decays, and more generally by high momentum charged particles is
used to uniquely assign an event to a tagging category.

Events are assigned a {\tt Lepton} tag
if they contain an identified lepton
with a center-of-mass momentum greater than
1.0 or 1.1\gevc\ for electrons and muons, respectively.
The momentum requirement selects mostly primary leptons by suppressing
opposite-sign leptons from semileptonic charm decays.
If the sum of charges of all identified kaons is non-zero,
the event is assigned a {\tt Kaon} tag.
The final two tags involve a multi-variable analysis based on a
neural network, which is trained
to identify primary leptons, kaons, and soft pions, and
the momentum and charge of the track with the maximum center-of-mass
momentum.  Depending on the output of the neural net, events are
assigned either an
{\tt NT1} (more certain)  tag, an {\tt NT2} (less certain) tag, or are 
considered not tagged (about 30\% of events) and excluded from the analysis.
The tagging power of the {\tt NT1} and {\tt NT2} tags comes primarily from
slow pions, from kinematically recovering non-identified primary
electrons and muons, and from kaons that do not pass the selection
criteria for the {\tt Kaon} category.

Tagging assignments are made mutually exclusive by the hierarchical
use of the tags. Events with a {\tt Lepton} tag and no conflicting {\tt Kaon}
tag are assigned to the {\tt Lepton} category. If no {\tt Lepton} tag exists,
but the event has a {\tt Kaon} tag, it is assigned to the {\tt Kaon} category.
Otherwise the event is assigned to one of the two neural network categories.

The effective tagging efficiency
$Q_i = \eps_i (1-2\mistag_i)^2$, where $\eps_i$ is the fraction of events assigned to
category $i$ and $\mistag_i$ the probability of obtaining a wrong tag, is used as the basis for optimization of
category selection criteria. The statistical errors on
\deltamd\ and \stwob\ are proportional to $1/\sqrt{Q}$, where $Q = \sum {Q_i}$.
The contributions of the various tagging categories to $Q$ is shown 
in Table \ref{tab:mistag}.

\begin{table}[thb]
\caption
{Tagging efficiency $\varepsilon$, average mistag fractions $\mistag$, mistag differences
  $\Delta\mistag=\mistag(\Bz)-\mistag(\Bzb)$, and the derived $Q$ (defined
in the text) obtained from the likelihood fit to the 
\Bflav\ and \Bcp\ samples.
%from the fully-reconstructed \Bz\ sample ($B_{\rm
%  flav}$ and $B_{\CP}$). 
%  Uncertainties are statistical only.  
  }
\label{tab:mistag}
\begin{tabular*}{\hsize}{|l
@{\extracolsep{0ptplus1fil}}  D{,}{\ \pm\ }{-1}
@{\extracolsep{10ptplus1fil}} D{,}{\ \pm\ }{-1}
@{\extracolsep{0ptplus1fil}}  D{,}{\ \pm\ }{-1}
@{\extracolsep{0ptplus1fil}}  D{,}{\ \pm\ }{-1}|}
\hline
Category     &
\multicolumn{1}{c}{$\ \ \ \varepsilon$   (\%)} &
\multicolumn{1}{c}{$\ \ \ \mistag$       (\%)} &
\multicolumn{1}{c}{$\ \ \ \Delta\mistag$ (\%)} &
\multicolumn{1}{c}{$\ \ \ Q$             (\%)} \\
\hline
\hline
{\tt Lepton} & 11.1,0.2 &  8.6, 0.9 &  0.6,1.5 &   7.6,0.4  \\
{\tt Kaon}   & 34.7,0.4 & 18.1, 0.7 & -0.9,1.1 &  14.1,0.6  \\
{\tt NT1}    & ~7.7,0.2 & 22.0, 1.5 &  1.4,2.3 &   2.4,0.3  \\
{\tt NT2}    & 14.0,0.3 & 37.3, 1.3 & -4.7,1.9 &   0.9,0.2  \\
\hline
All          & 67.5,0.5 &           &          &  25.1,0.8  \\
\hline
\end{tabular*}
\end{table}

\subsection{Mixing result}

%In the Standard Model,
%\Bz-\Bzb mixing occurs through second-order weak diagrams
%involving the exchange of up-type quarks,
%with the top quark contributing
%the dominant amplitude.  A measurement of  the difference between
%the mass eigenstates, \deltamd, is therefore
%sensitive to the value of the CKM matrix element
%$V_{td}$.
%At present the sensitivity to $V_{td}$ is not limited by
%experimental precision on \deltamd, but by other
%uncertainties in the calculation, in particular the quantity
%$f_B^2 B_B$, where $f_B$ is the \Bz\ decay constant, and
%$B_B$ is the so-called bag factor,
%representing  the strong interaction matrix elements.

The value of \deltamd\ is extracted from the tagged flavour-eigenstate
\Bz\ sample with a simultaneous unbinned likelihood fit to the
\deltat\ distributions of both unmixed and mixed events. 
The PDFs for the unmixed $(+)$ and mixed $(-)$
signal events for the $i^{th}$ tagging category are given by
\begin{equation}
\label{eq:mixing_likelihood}
{\cal H}_{\pm}(\deltat,\sigma_{\deltat}| \deltamd, \mistag_i, \hat{a}_i)  = \frac{{\rm e}^{ -\left| \deltat \right|/\tau}}{4\tau}
\left[ 1 \pm (1-2\mistag_i)\cos{ \deltamd \deltat } \right] \otimes {\cal R}(\delta_t,\sigma_{\deltat}|\hat {a}_i).
\end{equation}
Some resolution function parameters are allowed to differ for each
tagging category to account for shifts due to inclusion of charm decay
products in the tag vertex. The PDFs are extended to include background
terms, different for each tagging category. 
The probability that a \Bz\ candidate is a signal event is determined
from a fit to the observed \mes\ distribution for its tagging category.
The \deltat\ distributions of the combinatorial background are described
with a zero lifetime component and a non-oscillatory component with
non-zero lifetime. Separate resolution function parameters are used for
signal and background to minimize correlations.

\begin{figure}[!htb]
  \begin{center}
    \mbox{
    \includegraphics[width=0.421\linewidth]{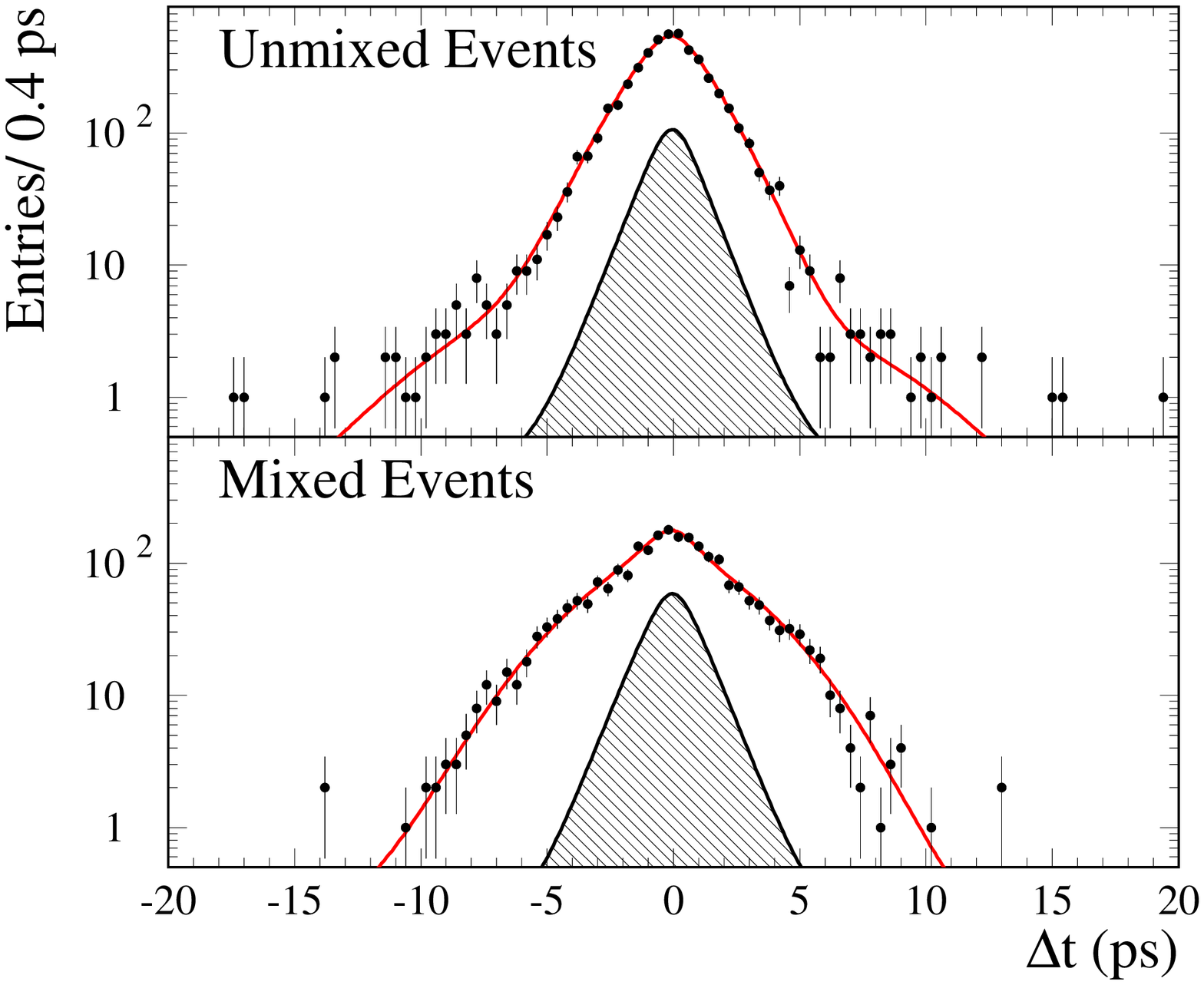}
    \includegraphics[width=0.569\linewidth]{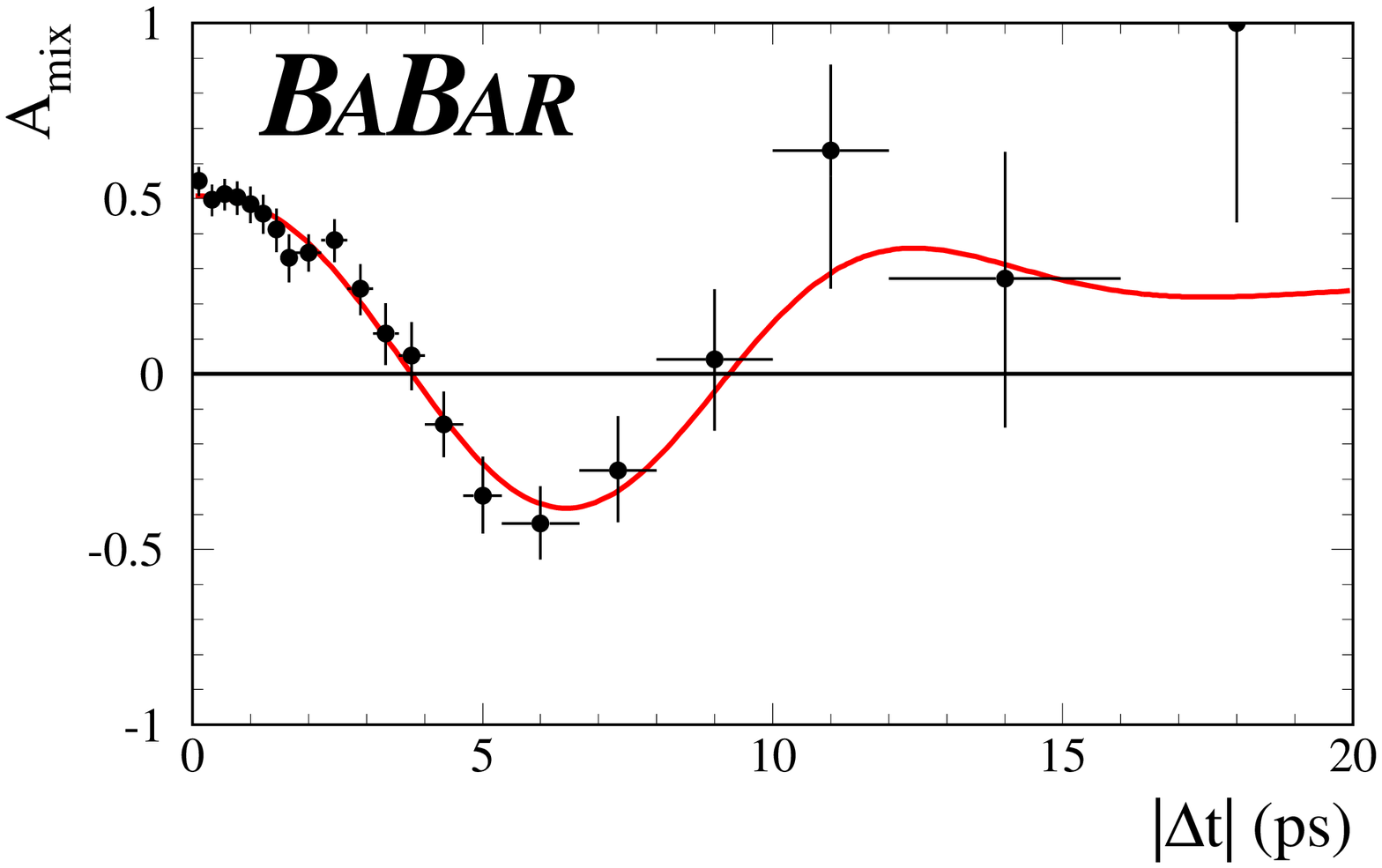}
\put(-285,140){{\large a)}}
\put(-285,70){{\large b)}}
\put(-25,140){{\large c)}}}
  \end{center}
\caption{Distributions of \deltat\ for (a) unmixed and
(b) mixed events in the signal region $\mes > 5.27$\gevcc.
The data points are overlaid with the result from the 
fit, projected using the individual
signal probabilities and event-by-event \deltat\ resolutions,
%The \deltat\ distributions obtained from the likelihood fit
%are overlaid, 
along with the simultaneously determined
background distribution. % shown as the curve in b).
Also shown (c) is the time-dependent mixing asymmetry ${\cal A}_{mix}(|\deltat|)$ 
defined in the text.
   \label{fig:b0.excl-data.deltat.all}}
\end{figure}

The $\Delta t$ distributions of the signal ($\mes>5.27\gevcc$) 
%and background ($\mes<5.27\gevcc)$ candidates, 
overlaid with the projections
of the likelihood fit, are shown in Fig.~\ref{fig:b0.excl-data.deltat.all}.
In addition, the mixing asymmetry,
\begin{equation}
{\cal A}_{mix}(\deltat) =
\frac{N_{unmixed}(\deltat)-N_{mixed}(\deltat)}{N_{unmixed}(\deltat)+N_{mixed}(\deltat)}. 
\label{eq:asym}
\end{equation}
is plotted.
If flavour tagging and \deltat\ determination were perfect, the asymmetry
as a function of \deltat\ would be a cosine with unit amplitude.
%The observed amplitude is diluted by mistag probabilities and the 
%finite \deltat\ resolution.

%The results from the likelihood fit to the tagged \Bz\ sample are
%summarized in Table~\ref{tab:result-likeli}.  
The probability
to obtain a likelihood smaller than that observed is 44\%, evaluated with a
parameterized Monte Carlo technique.  The value of
\deltamd\ obtained is
\begin{equation}
\deltamd = 0.516 \pm 0.016 ({stat}) \pm 0.010 ({syst})\,\hips.
\end{equation}
Since the parameters of the \deltat\ resolution and the mistag rates \mistag % for both signal and backgrounds
are free parameters in the fit, their contribution to the uncertainty on
\deltamd\ is included as part of the statistical error.
The main contributions to the systematic errors are the choice
of the signal \deltat\ resolution description, its capability to
handle outliers and various worst-case SVT misalignment scenarios
($\pm 0.005\hips$), and by correlations between mistag rates
and \deltat\ resolution which are not explicitly modeled by the
likelihood fit ($\pm 0.005\hips$).
Finally, the variation of the fixed
\Bz\ lifetime within known errors~\cite{PDG2000} leads to
a systematic uncertainty of $\pm 0.006\hips$.

This is one of the single most precise mixing measurements available\cite{mixing-prl}, and
is consistent with the current world average~\cite{PDG2000}.
%and a recent \babar\ measurement with a dilepton sample~\cite{dilep-mixing-prl},
%which obtains $\deltamd = 0.493 \pm 0.012 ({stat}) \pm 0.009 ({syst})\,\hips$.
%These two measurements, together with the Belle dilepton measurement\cite{belle-dilep-mixing-prl} of
%$\deltamd = 0.463 \pm 0.008 ({stat}) \pm 0.016 ({syst})\,\hips$, have resulted in an
%updated world average for \deltamd\ which is almost twice as precise as 
%the current one.

\section{Determination of \stwob}

\begin{figure}[!h]%1
\begin{center}%
\mbox{\epsfig{figure=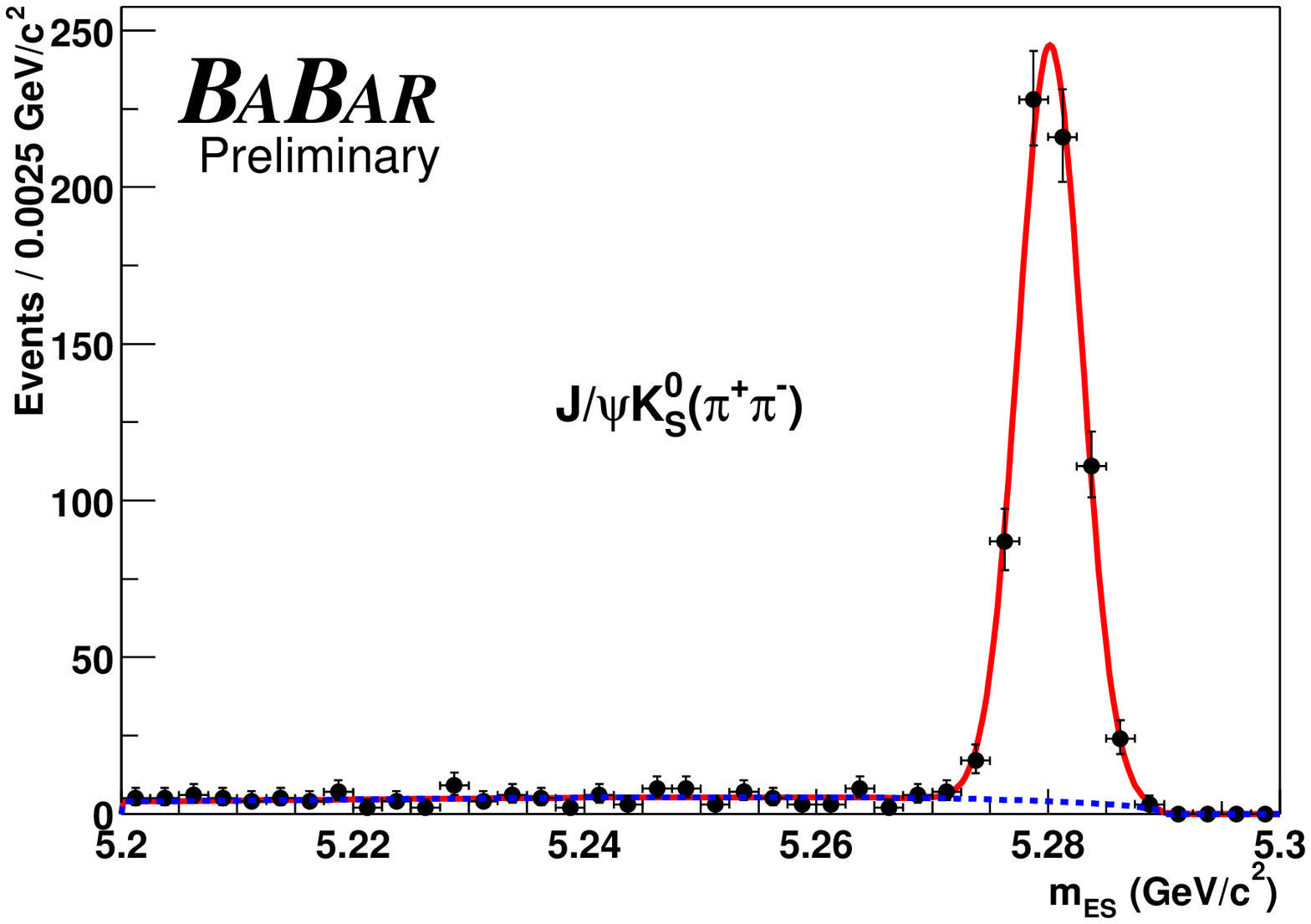,width=0.45\linewidth,clip=}
\epsfig{figure=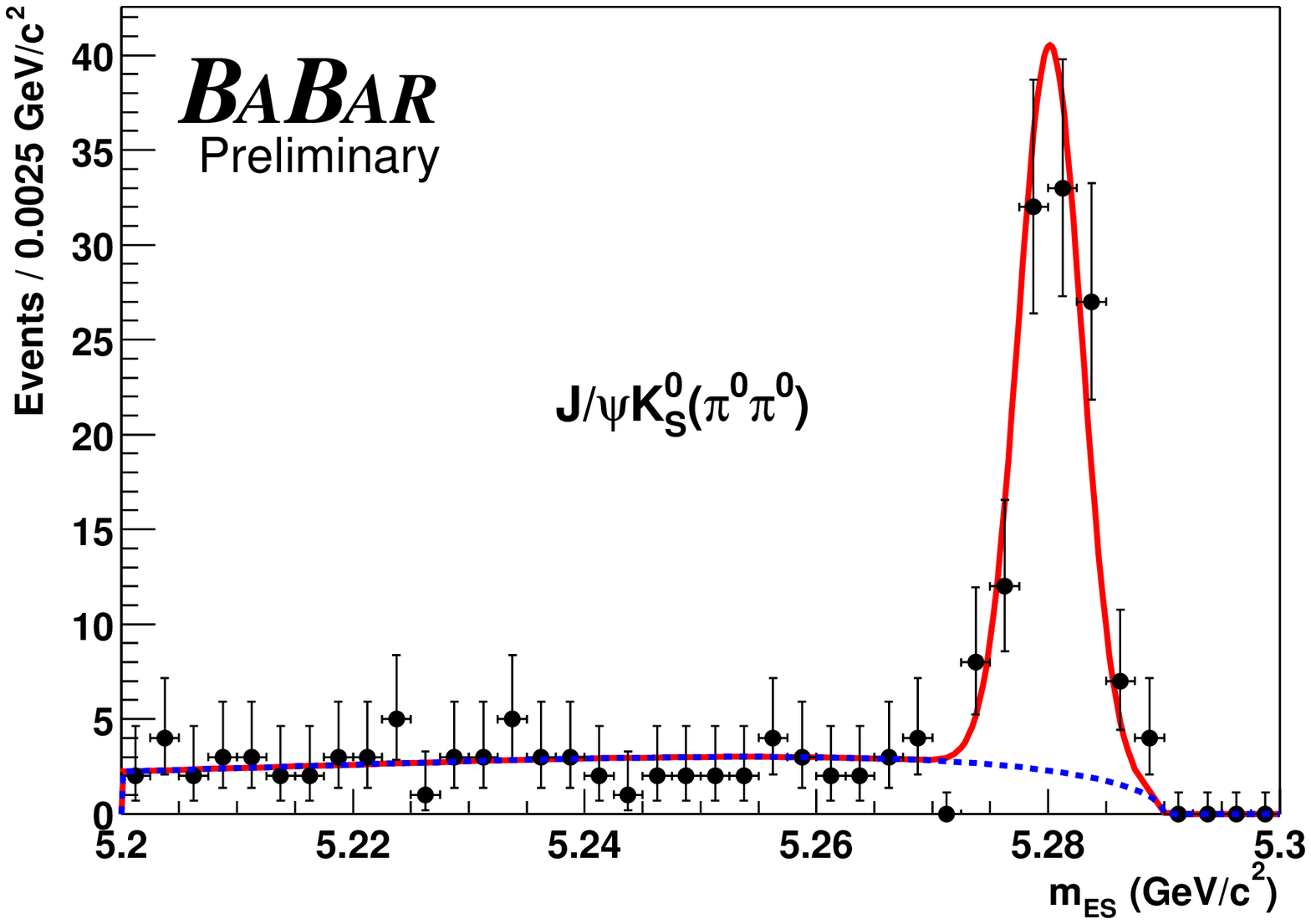,width=0.45\linewidth,clip=}
\put(-245,120){{\large a)}}
\put(-37,120){{\large b)}}}
\mbox{\epsfig{figure=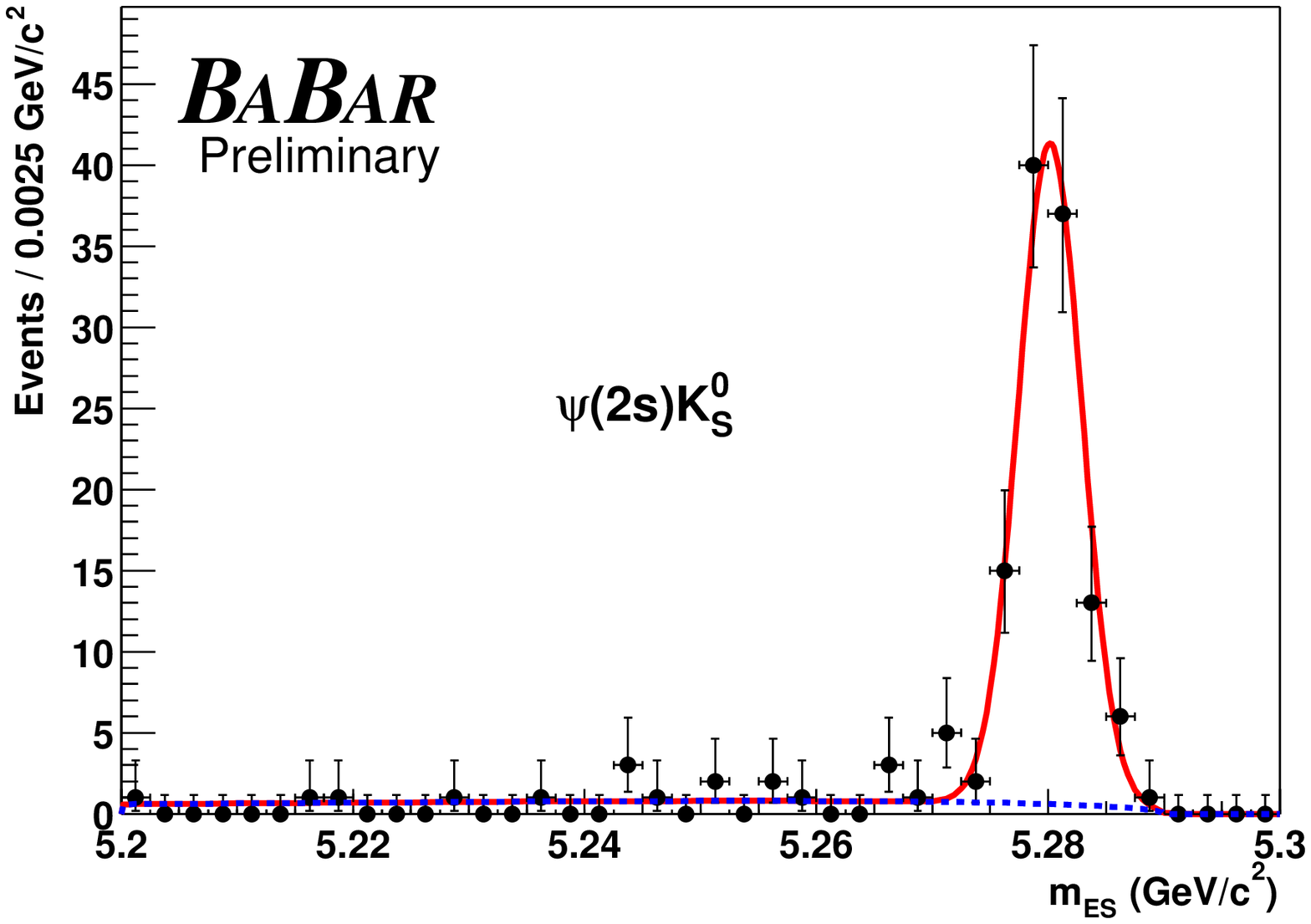,width=0.45\linewidth,clip=}
\epsfig{figure=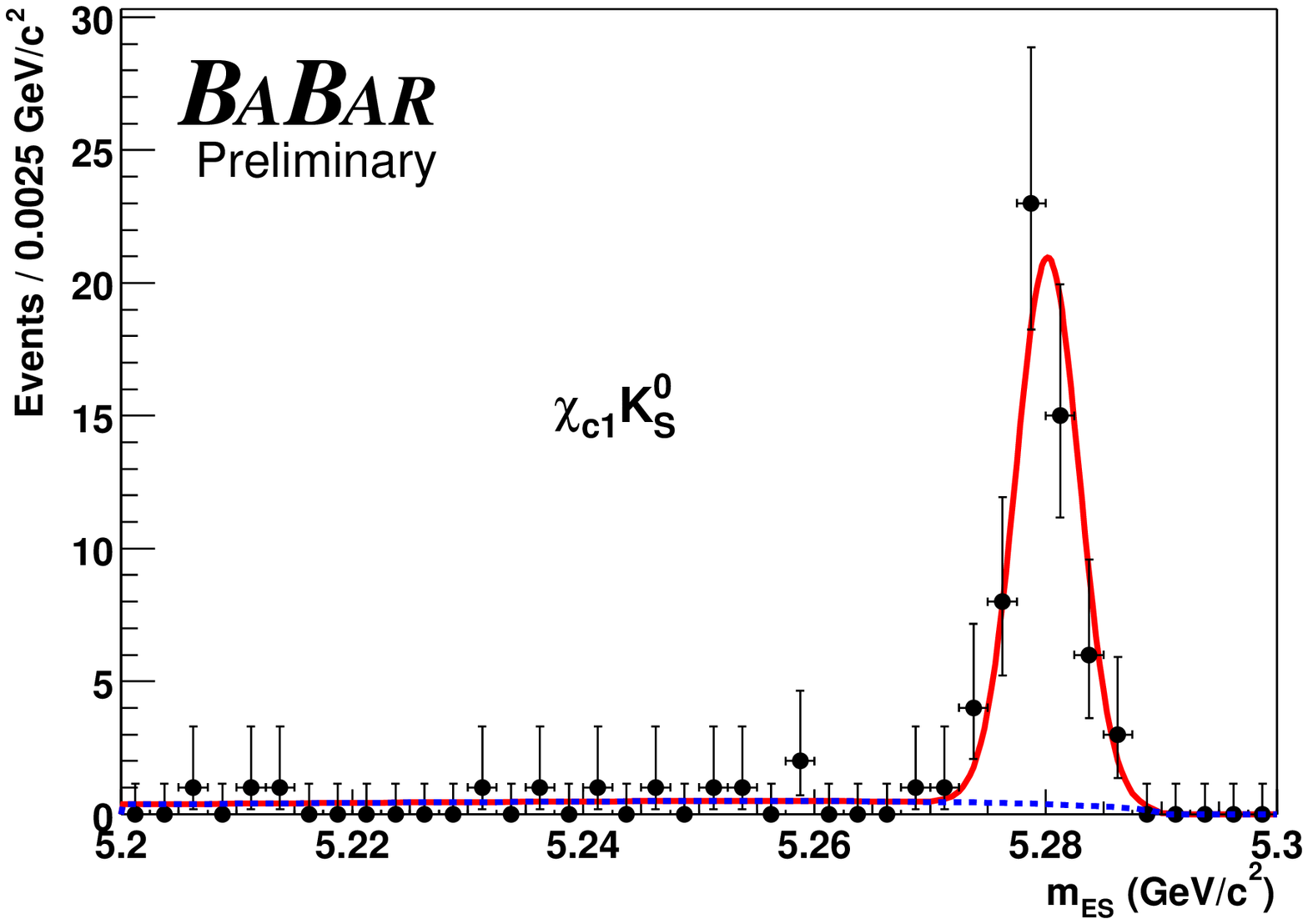,width=0.45\linewidth,clip=}
\put(-245,120){{\large c)}}
\put(-37,120){{\large d)}}}
\mbox{\epsfig{figure=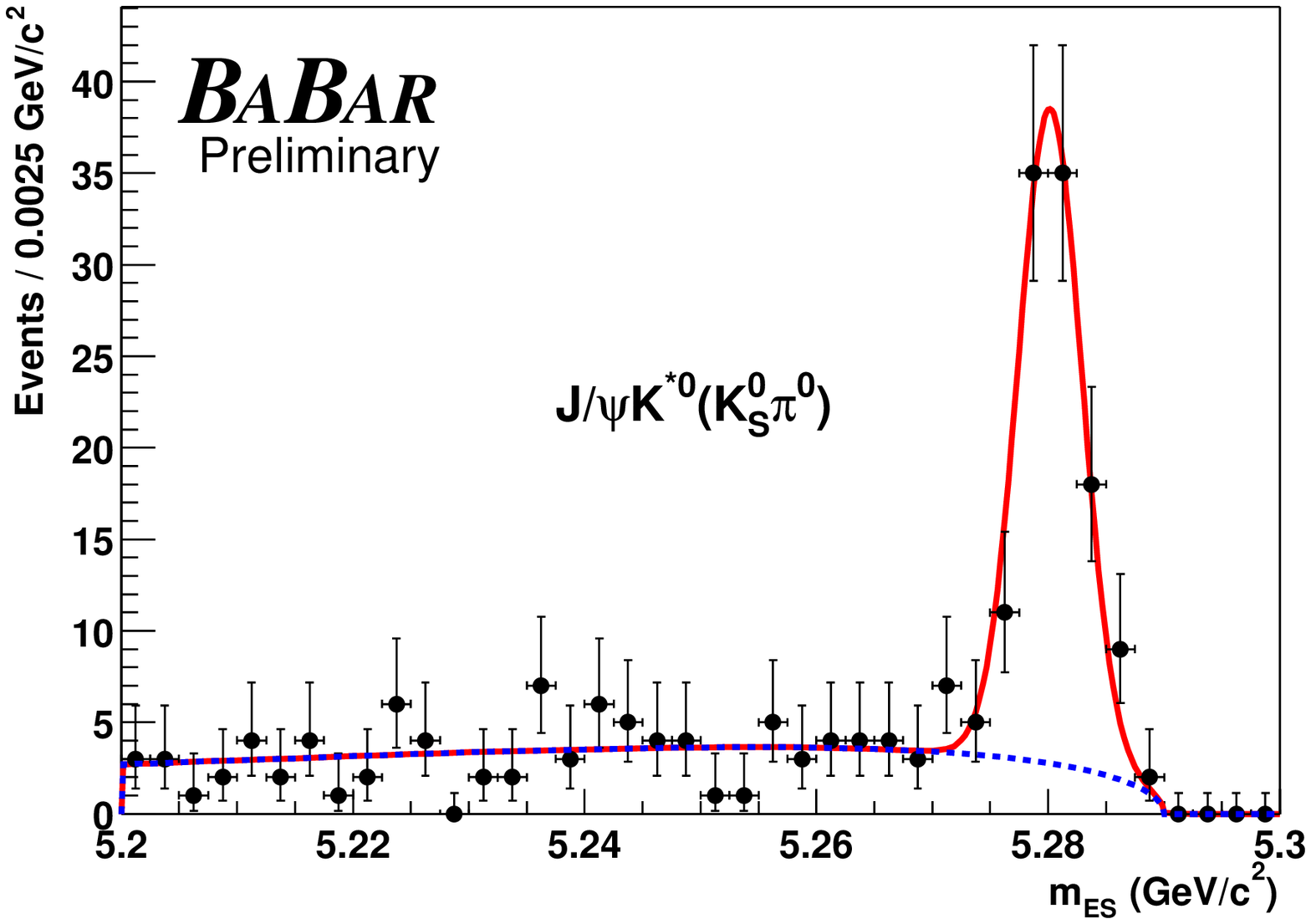,width=0.45\linewidth,clip=}
\epsfig{figure=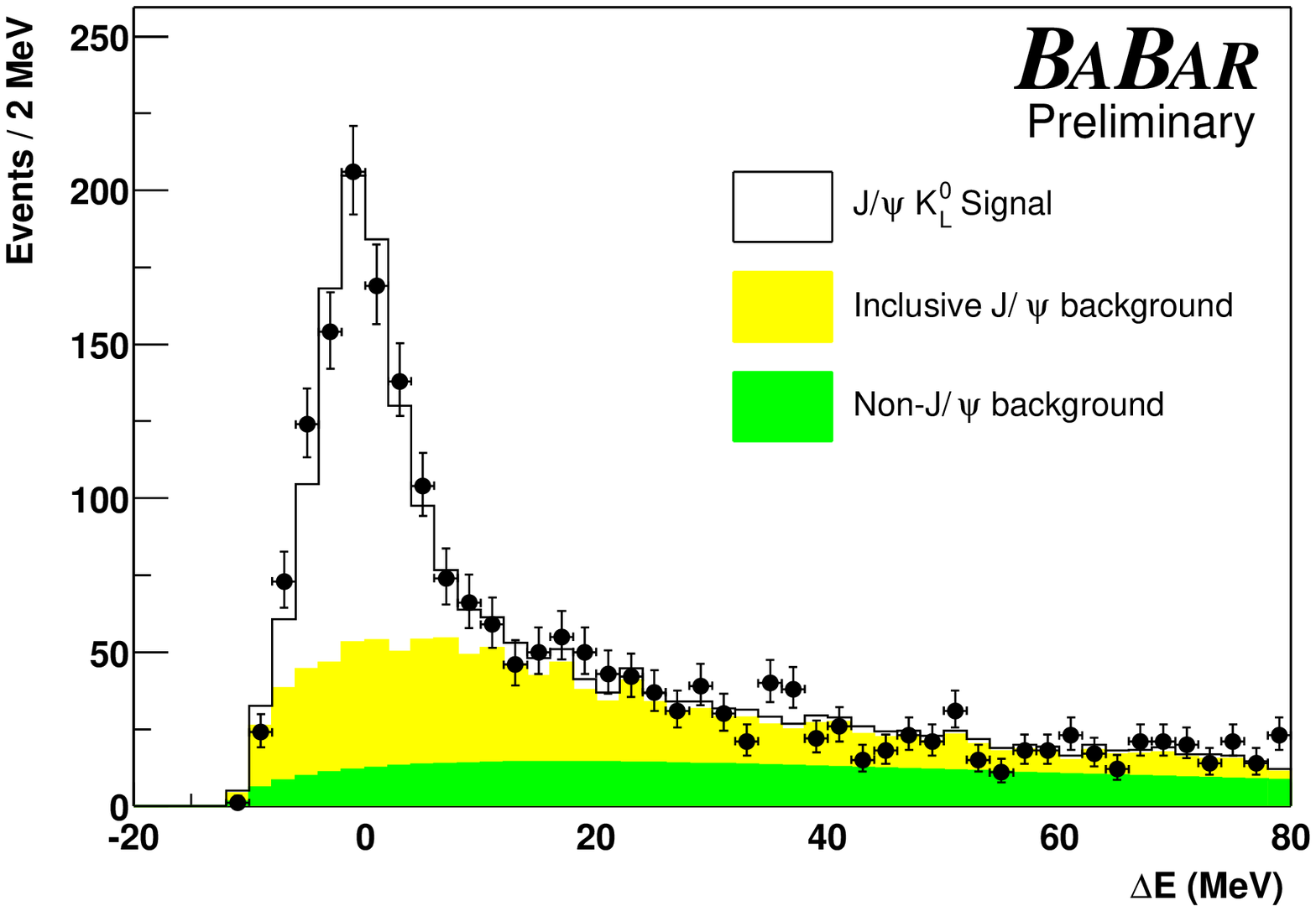,width=0.45\linewidth,clip=}
\put(-245,115){{\large e)}}
\put(-175,115){{\large f)}}}
\caption{Distribution of \mes\ for flavour tagged $B_{\CP}$ candidates
  selected in the final states
  a) $J/\psi K^0_S\ (\KS\to\pip\pim)$,
  b) $J/\psi K^0_S\ (\KS\to\piz\piz)$,
  c) $\psi(2S) K^0_S$,
  d) $\chi_{c1} K^0_S$,
  e) $J/\psi K^{*0}(K^{*0}\to K^0_S\pi^0)$, and
  f) distribution of $\Delta E$ for flavour tagged $\jpsi\KL$
  candidates.}
\label{fig:bcpsample}
\end{center}
\end{figure}

For the measurement of \stwob, \Brec\ is fully
reconstructed in a
\CP\ eigenstate with eigenvalue $\etaCP=-1$ ($\jpsi \KS$, $\psitwos \KS$, or
$\chicone\KS$)
or $+1$ ($\jpsi\KL$), while \Btag\
is tagged just as for the mixing measurement.
The sample is further
enlarged by including the mode $\jpsi\Kstarz$ ($\Kstarz\to\KS\piz$). However,
due to the presence of even ($L=0$, 2) and odd ($L=1$)
orbital angular momenta in the $\jpsi\Kstarz$ system,
there are $\etaCP=-1$ and $+1$ contributions to its decay rate. 
These contributions are disentangled by incorporating 
their dependence on the transversity angles in each event into
the likelihood fit\cite{sin2b-conf-paper}. 
The \mes distributions (\deltaE\ for \jpsi\KL) of the selected 
sample
are shown in Fig.~\ref{fig:bcpsample}, and the detailed breakdown
in Table~\ref{tab:result}.

\begin{table}[!thb]
\caption{
Number of tagged events, signal purity, and result of fitting for \CP
asymmetries in the full \CP sample and in various subsamples, as well as
in the $B_{\rm flav}$ and charged $B$ control samples. Purity is the
fitted number of signal events divided by the total number of events in
the $\Delta E$ and \mes signal region defined in the text.
%Errors are statistical only.
}
\label{tab:result}
\begin{tabular*}{\hsize}{| l@{\extracolsep{0ptplus1fil}} r c@{\extracolsep{0ptplus1fil}} D{,}{\ \pm\ }{-1} |}
\hline
 Sample  & $N_{\rm tag}$ & Purity (\%) & \multicolumn{1}{c}{$\ \ \ \stwob$}
\\ \hline
 Full \CP\ sample                        & 1850  & 79 &  0.75, 0.09 \\
\hline
$\ \jpsi \KS$ ($\KS \to \pi^+ \pi^-$)    & 693  & 96  &  0.79, 0.11 \\
$\ \jpsi \KS$ ($\KS \to \pi^0 \pi^0$)    & 123  & 89  &  0.42, 0.33 \\
$\ \psi(2S) \KS$                         & 119  & 89  &  0.84, 0.32 \\
$\ \chicone \KS $                        &  60  & 94  &  0.84, 0.49 \\
$\ \jpsi \KL$                            & 742  & 57  &  0.73, 0.19 \\
$\ \jpsi\Kstarz$ ($\Kstarz \to \KS\piz$) & 113  & 83  &  0.62, 0.56 \\
\hline
\hline
$\jpsi\KS$, $\psitwos\KS$, $\chicone\KS$ only $(\eta_f=-1)$ &  995  & 94 &  0.76, 0.10 \\
\hline
$\ $ {\tt Lepton} tags                   & 176  & 97  &  0.73, 0.16 \\
$\ $ {\tt Kaon} tags                     & 504  & 95  &  0.75, 0.14 \\
$\ $ {\tt NT1} tags                      & 117  & 95  &  0.86, 0.33 \\
$\ $ {\tt NT2} tags                      & 198  & 94  &  0.84, 0.61 \\
\hline
$\ $ \Bz\ tags                           & 471  & 94  &  0.79, 0.14 \\
$\ $ \Bzb\ tags                          & 524  & 95  &  0.73, 0.14 \\
\hline\hline
$B_{\rm flav}$ sample                    & 17546 & 85 &  0.00, 0.03 \\
Charged $B$ sample                       & 14768 & 89 & -0.02, 0.03 \\
\hline
\end{tabular*}
\end{table}

The decay-time distribution of $B$ decays to a \CP eigenstate with a \Bz
or \Bzb tag can be expressed in terms of a complex parameter $\lambda$
that depends on both the \Bz-\Bzb oscillation amplitude and the amplitudes
describing \Bzb and \Bz decays to this final
state~\cite{lambda}. The decay rate  ${\cal F}_+({\cal F}_-)$ when the
tagging meson is a $\Bz (\Bzb)$ is given by
\begin{eqnarray}
{\cal F}_\pm(\deltat,\sigma_{\deltat}|\stwob,\mistag_i,\hat {a}_i) \hspace*{-2.8cm}&& \nonumber \\ &=&  
%\frac{\Gamma\, {\rm e}^{ - \Gamma | \deltat | }}{ 2(1  + |\lambda|^2)}  
%\left\{ \frac{1}{2}\left(1  + |\lambda|^2\right) 
%\pm {\left(1-2\mistag\right)} \left[
%Im(\lambda) \sin{ \deltamd  \deltat } 
%-\frac{1}{2}\left(1-|\lambda|^2\right) \cos{ \deltamd  \deltat }  \right]  \right\}\\
{\frac{{\rm e}^{{- \left| \deltat \right|}/\tau_{\Bz} }}{4\tau_{\Bz}
}}   \left[ \ 1 \hbox to 0cm{}
\pm \left(1-2\mistag_i\right)\left( \frac{{2\mathop{\cal I\mkern -2.0mu\mit m}}
\lambda}{1+|\lambda|^2}  \sin{( \deltamd  \deltat )} 
- { \frac{1  - |\lambda|^2 } {1+|\lambda|^2} }  
  \cos{( \deltamd  \deltat) } \right)  \right] \nonumber \\
&\otimes& {\cal R}(\delta_t,\sigma_{\deltat}|\hat {a}_i),
\label{eq:direct}
\end{eqnarray}
where the $+$ or $-$ sign indicates whether the \Btag\ is
tagged as a \Bz\ or a \Bzb, respectively. 

The distributions are much simpler when $|\lambda|=1$, which is the
expectation of the Standard Model for decays like $\Bz\to \jpsi\KS$
where all amplitudes which contribute to the decay have the 
same weak phase.  In this particular case one is left with the phase 
difference introduced by \Bz-\Bzb\ mixing, i.e. $\lambda=\etaCP {\rm e}^{2i\beta}$.
where $\etaCP$ is the \CP\ eigenvalue of the final state.  
%The remaining factor is the phase
%introduced by \Bz-\Bzb\ mixing.

It is possible to construct a \CP-violating observable
which, neglecting resolution effects, is proportional to \stwob:
\begin{equation}
  {\cal A}_{CP}(\deltat) = \frac{ {\cal F}_+(\deltat) - {\cal F}_-(\deltat) }
{ {\cal F}_+(\deltat) + {\cal F}_-(\deltat) } 
  \propto -\etaCP {\left( 1 - 2 \mistag\right)} \sin{ 2 \beta } \sin{ \deltamd \deltat }.
\label{eq:asymmetry}
\end{equation}
Since no time-integrated \CP\ asymmetry effect is expected, an
analysis of the time-dependent asymmetry is necessary.
The interference between the
two amplitudes, and hence the \CP\ asymmetry, is maximal
after approximately 2.1 \Bz\ proper lifetimes, when the mixing asymmetry
goes through zero. However, the maximum sensitivity to \stwob,
which is proportional to ${\rm e}^{-\Gamma \deltat}\sin^2\deltamd\deltat$, occurs
in the region of 1.4 lifetimes.

The value of \stwob\ can be extracted 
by maximizing the likelihood function
\begin{equation}
\label{eq:Likelihood}
 \ln { {\cal {L} }_{CP} } = \sum_{i}^{\rm tagging}
 \left[ \sum_{\Bz\ {\rm tag} } {  \ln{ {\cal F}_+( \deltat,\sigma_{\deltat}; \stwob, \mistag_i, \hat {a}_i, 
 ) } } + 
 \sum_{\Bzb\ {\rm tag} }{ \ln{ {\cal F}_-( \deltat,\sigma_{\deltat}; \stwob, \mistag_i, \hat {a}_i, 
 ) } } \right] ,
\end{equation}
where the outer summation is over tagging categories $i$ and
the inner summations are over the \Bz\ and \Bzb\ tags within a given
tagging category.
In practice,
the fit for \stwob\ is performed
on the combined \Bflav\ and \Bcp\ samples with a likelihood constructed
from the sum of Eq.~\ref{eq:mixing_likelihood} and \ref{eq:Likelihood},
in order to determine \stwob, the mistag fraction
$\mistag_i$ for each tagging category, and the resolution
parameters $\hat {a}_i$ simultaneously. Additional terms are included in the likelihood
to account for backgrounds and their time dependence.
The determination of the mistag fractions and \deltat resolution
function for the signal is dominated by the high-statistics 
\Bflav\ sample.  We fix $\tau_{\Bz}=1.548\ps$ and $\deltamd
 =0.472\ps^{-1}$~\cite{PDG2000}. 
The largest correlation between \stwob\ and any linear combination of 
other free parameters is only 0.14.

%\begin{figure}[htp]
%\begin{center}
%\epsfig{figure=figures/breco-control-dt-and-asym.eps,width=0.89\linewidth}
%\put(-310,295){{\large a)}}
%\put(-310,200){{\large b)}}
%\put(-310,120){{\large c)}}
%\caption{Number of $B_{\rm flav}$ candidates
%in the signal region a) with a \Bz tag, $N_{\Bz }$, and b)
%with a \Bzb tag, $N_{\Bzb}$, and c) the raw asymmetry
%$(N_{\Bz}-N_{\Bzb})/(N_{\Bz}+N_{\Bzb})$, as functions of \deltat . The
%solid curves represent the result of the combined fit to all selected
%$B_{\rm flav}$ events. The shaded regions represent the background
%contributions.}
%\label{fig:bflavasym}
%\end{center}
%\end{figure}

\begin{figure}[tp]
\begin{center}
\epsfig{figure=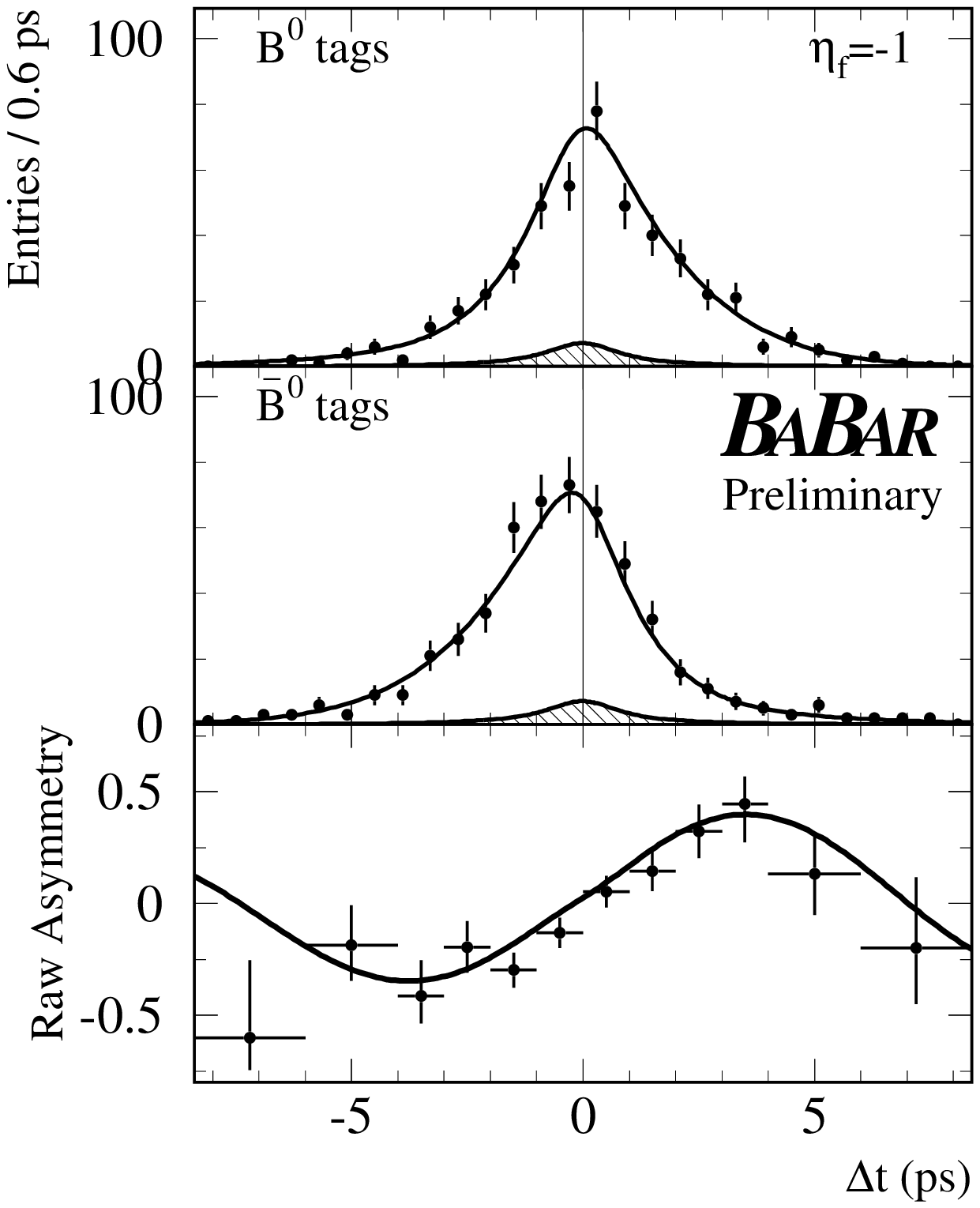,width=0.49\linewidth}
\epsfig{figure=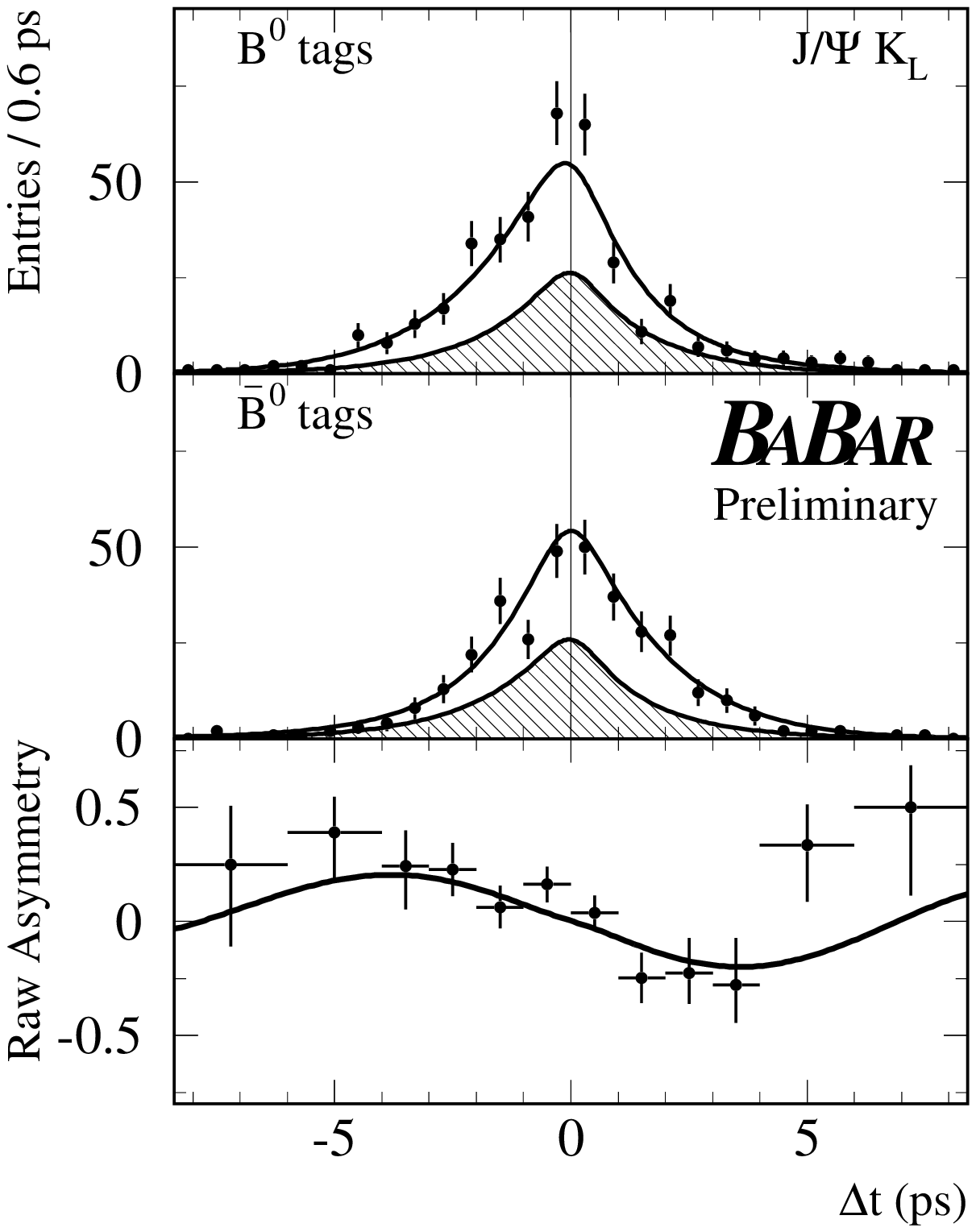,width=0.49\linewidth}
\put(-408,270){{\large a)}}
\put(-182,270){{\large d)}}
\put(-408,188){{\large b)}}
\put(-182,188){{\large e)}}
\put(-408,105){{\large c)}}
\put(-182,105){{\large f)}}
\caption{Number of $\eta_f=-1$ candidates
($J/\psi \KS$,
$\psi(2S) \KS$,
$\chicone \KS$)
in the signal region a) with a \Bz tag $N_{\Bz }$ and b)
with a \Bzb tag $N_{\Bzb}$, and c) the raw asymmetry
$(N_{\Bz}-N_{\Bzb})/(N_{\Bz}+N_{\Bzb})$, as functions of \deltat . The
solid curves represent the result of the combined fit
to the full $B_{\rm CP}$ sample.
%to all selected \CP events. 
The shaded regions represent the background contributions.
Figures d) -- f) contain the corresponding information for the $\eta_f=+1$
mode $J/\psi \KL$.
The likelihood is normalized to the total number of \Bz and \Bzb tags.
The value of \stwob is independent of the individual
normalizations and therefore of the difference between the number of \Bz
and \Bzb tags. This difference is responsible for the small vertical
shift between the data points and the solid curves.}
\label{fig:cpdeltat}
\end{center}
\end{figure}
The simultaneous fit to all \CP\ decay modes and the flavour decay
modes yields:
\be
    \stwob = 0.75 \pm 0.09 \stat \pm 0.04 \syst.
\ee
The dominant sources of systematic uncertainties are the choice of
parameterization of the \deltat\ resolution function, possible
differences in the mistag fractions between the \CP\ sample and the
flavour sample, and uncertainties in the level, composition and \CP\ 
asymmetry of the background in the selected events. 
The large sample of fully reconstructed events allows a number of consistency
checks, including separation of the data by decay mode, tagging category
and \Btag\ flavour.

This analysis\cite{sin2b-conf-paper} improves upon and supercedes the previously published result\cite{sin2b-prl}.
It provides the single most precise measurement of \stwob\ currently available
and is consistent with the range implied by indirect measurements and theoretical estimates
of the magnitudes of CKM matrix elements in the context of the Standard Model\cite{CKMconstraints}.

%\section*{Acknowledgments}
%This is where one places acknowledgments for funding bodies etc.
%Note that there are no section numbers for the Acknowledgments, Appendix
%or References.

%\section*{Appendix}


\section*{References}
\begin{thebibliography}{99}

\bibitem{EpsilonK}
J.H.~Christenson {\em et al.}, \jprl {13}, 138 (1964); 
NA31 Collaboration, G.D.~Barr {\em et al.}, \jpl {317}, 233 (1993);
E731 Collaboration, L.K.~Gibbons {\em et al.}, \jprl {70}, 1203 (1993).

\bibitem{CKM}
N.~Cabibbo, \jprl {\bf 10}, 531 (1963);
M.~Kobayashi and T.~Maskawa, \progtp {\bf 49}, 652 (1973).

\bibitem{MSConstraints}
See, for instance, ``Overall determinations of the CKM matrix'',
Section~14 in ``The \babar\ physics book'', P.\ H.\ Harrison and H.\ R.\ Quinn, eds.,
SLAC-R-504 (1998), and references therein.

\bibitem{Primer}
For an introduction to \CP\ violation, see, for instance,
``A \CP\ violation primer'', Section~1 in ``The \babar\ physics book'',
op. cit.~\cite{MSConstraints},
and references therein.

\bibitem{pepii}
``PEP-II: An Asymmetric $B$ Factory'',
Conceptual Design Report, SLAC-418, LBL-5379 (1993).

\bibitem{sin2b-prl}
\babar\ Collaboration, B.\ Aubert {\em et al.},
\PRL {\bf 87}, 091801;
\babar\ Collaboration, B.\ Aubert {\em et al.},
\babar -PUB-01/03, SLAC-PUB-9060, hep-ex/0201020, to appear in \PRD.

\bibitem{babar-detector-nim}
\babar\ Collaboration, B.\ Aubert {\em et al.},
\nim{\bf A479}, 117 (2002).


\bibitem{geant4}
{\tt http://wwwinfo.cern.ch/asd/geant4/geant4.html}

\bibitem{fox}
G.C.~Fox and S.~Wolfram, \jprl{41}, 1581 (1978).

\bibitem{blife-prl}
\babar\ Collaboration, B.\ Aubert {\em et al.},
\PRL {\bf 87}, 201803

\bibitem{PDG2000}
Particle Data Group, D.E.~Groom {\em et al.}, \epjc{15}, 1 (2000).

\bibitem{mixing-prl}
\babar\ Collaboration, B.\ Aubert {\em et al.}, 
\babar -PUB-01/02, SLAC-PUB-9061, hep-ex/0112044, to appear in \PRL.

%\bibitem{dilep-mixing-prl}
%\babar\ Collaboration, B.\ Aubert {\em et al.}, 
%\babar -PUB-01/22, SLAC-PUB-9096, hep-ex/0112045, to appear in \PRL.

%\bibitem{belle-dilep-mixing-prl}
%Belle Collaboration, K.\ Abe {\em et al.},
%\PRL {\bf 86}, 3228

\bibitem{sin2b-conf-paper}
\babar\ Collaboration, B.\ Aubert {\em et al.},
\babar -CONF-02/01, SLAC-PUB-9153, hep-ex/0203007.

\bibitem{lambda}
See, for example, L.~Wolfenstein, \epjc{15}, 115 (2000).


\bibitem{CKMconstraints}
See, for example, F.J.~Gilman, K.~Kleinknecht and B.~Renk,
\epjc{15}, 110 (2000).


\end{thebibliography}
\end{document}